\documentclass[conference]{IEEEtran}
\IEEEoverridecommandlockouts

\usepackage{cite}
\usepackage{amsmath,amssymb,amsfonts}
\usepackage{algorithmic}
\usepackage{graphicx}
\usepackage{textcomp}
\usepackage{xcolor}
\usepackage{url}
\def\BibTeX{{\rm B\kern-.05em{\sc i\kern-.025em b}\kern-.08em
    T\kern-.1667em\lower.7ex\hbox{E}\kern-.125emX}}
\usepackage{hyperref}
    
\usepackage{booktabs}

\begin{document}

\title{Deep Learning Based EDM Subgenre Classification using Mel-Spectrogram and Tempogram Features}

\author{\IEEEauthorblockN{Wei-Han Hsu, Bo-Yu Chen, and Yi-Hsuan Yang}
\IEEEauthorblockA{\textit{Research Center for IT Innovation} \\
\textit{Academia Sinica}\\
Taipei, Taiwan \\
\{ddmanddman,~bernie40916,~yang\}@citi.sinica.edu.tw}
}

\maketitle

\begin{abstract}
Along with the evolution of music technology, a large number of styles, or ``subgenres,'' of Electronic Dance Music (EDM) have emerged in recent years. While the classification task of distinguishing between EDM and non-EDM has been often studied in the context of music genre classification,  little work has been done on the more challenging EDM subgenre classification. The state-of-art model is based on extremely randomized trees and could be improved by deep learning methods. In this paper, we extend the state-of-art music auto-tagging model ``short-chunk CNN$+$Resnet'' to EDM subgenre classification, with the addition of two mid-level tempo-related feature representations, called the Fourier tempogram and autocorrelation tempogram. 
And, we explore two fusion strategies, early fusion and late fusion, to aggregate the two types of tempograms. 
We evaluate the proposed models using a large dataset consisting of 75,000 songs for 30 different EDM subgenres, and show that the adoption of deep learning models and tempo features indeed leads to higher classification accuracy. 

\end{abstract}

\begin{IEEEkeywords}
EDM subgenre classification, deep learning, feature fusion, tempogram, convolutional neural network
\end{IEEEkeywords}

\section{Introduction}

Electronic Dance Music (EDM) is a kind of dance and club music. Disk Jocket (DJs) usually need to classify EDM by their subgenres to get songs with similar style for many purposes, such as for making the DJ set transitions part.
An automatic program for EDM subgenre classification can be useful for human DJs \cite{sankalp14}. It is also a fundamental building block towards realizing an automatic AI DJ \cite{djnet,djtransgan}.  

The task \emph{automatic EDM subgenre classification} can be in general considered as an instance of music auto-tagging problem \cite{spm}.
Therefore, methodology-wise we can base on research that has been done for general music auto-tagging and classification. 
However, we note that, due to the similarity among the EDM subgenres, EDM subgenre classification can sometimes be difficult even for human DJs. 
The decision between genres can be fuzzy.
For example,  ``tech-house,'' ``deep-house,'' and ``progressive-house'' music may sound fairly similar as they are all ``house'' music.

We present in this paper a deep learning based approach to automatic EDM subgenre classification, extending the recent work by Caparrini \emph{et al.}~\cite{edm}, which use a non-deep learning approach.
Following their work, we compile our training and test sets from Beatport (\url{https://www.beatport.com/}), a worldwide principal source of music for DJs.
The Beatport website assigns only a single subgenre label to each song, so we can formulate the task as a multi-class classification problem. Our dataset contains 30 different subgenres, each with 2,500 songs and hence 75,000 songs in total. We thus treat it as a 30-class classification problem.

\begin{figure}
    \centering
    \includegraphics[width=0.484\textwidth]{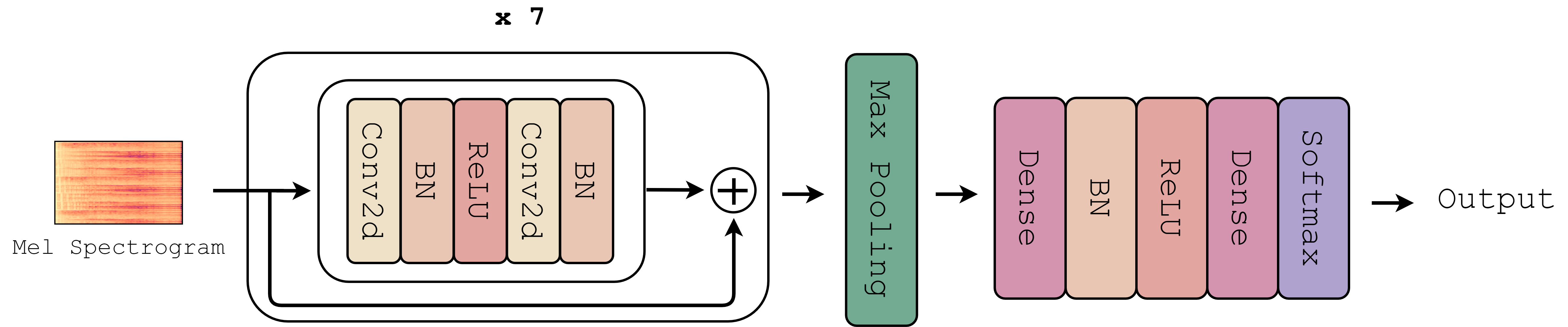}
    \caption{System diagram of the short-chunk CNN$+$Resnet model \cite{sota}.}
    \label{fig:short-chunk cnn}
\end{figure}

While the classifier used by Caparrini \emph{et al.}~\cite{edm} was based on extremely randomized trees~\cite{ert}, a non-deep learning algorithm, our classifier is based on the ``short-chunk convolutional neural network (CNN)$+$Resnet'' deep architecture proposed by Won \emph{et al.}~\cite{sota}, which represents the state-of-the-art in music auto-tagging (see Figure \ref{fig:short-chunk cnn}).
However, the original short-chunk CNN model takes only the Mel-spectrograms as input for feature learning.
While this may be sufficient for classifying broader genre classes such as Pop, Rock, Jazz, and EDM,  it is unclear how it performs for a subgenre classification task.

\begin{figure}
    \centering
        \includegraphics[width=0.484\textwidth]{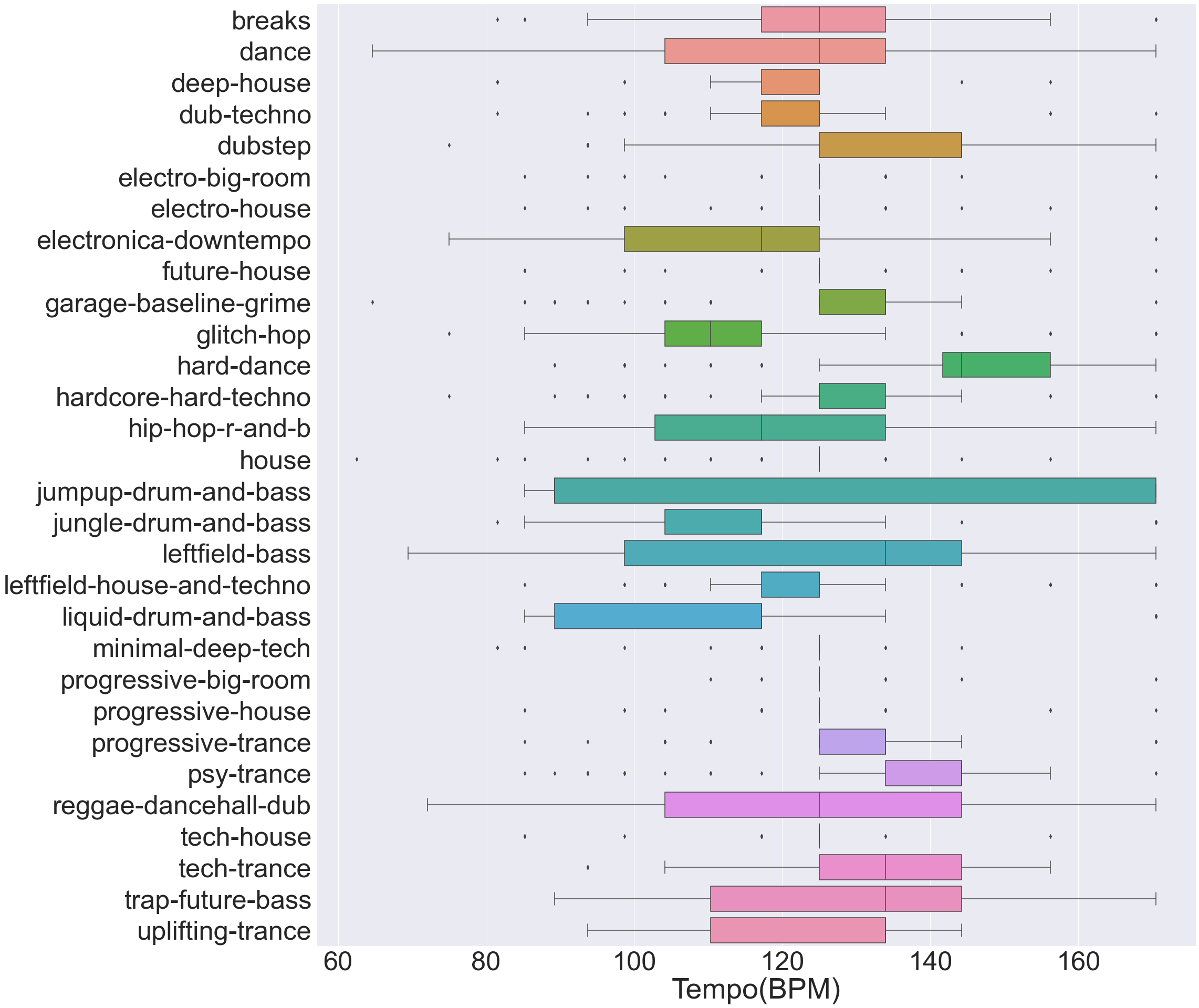}
    \caption{Boxplots of the tempo values (in beat-per-minute; or BPM) of different EDM subgenres considered in our work.}
    \label{fig:boxplot}
\end{figure}

In particular, we note that, among the 92 hand-crafted audio features employed by  Caparrini \emph{et al.}~\cite{edm}, 
\emph{tempo}-related features were found to be the top-four important features. 
By estimating the tempo values for all the songs in our dataset using the algorithm of Grosche \emph{et al.}~\cite{tempogram}, and plotting the resulting distribution of tempo values for 100 randomly picked songs from each genre, as shown in Figure \ref{fig:boxplot}, we can see that different EDM subgenres do prefer different tempo values. 
For example, the tempo of ``psy-trance'' music is usually around 140 beat-per-minute (BPM), while ``deep-house'' is usually around 120 BPM. 
Different subgenres also have different tempo ranges. 
For example, the tempo of ``jumpup-drum-and-bass''  can vary widely from 90 to 170 BPM, while the tempo values of ``tech-house'' typically fall within 120 to 130 BPM.

In light of this, we propose simple extensions of the short-chunk CNN$+$Resnet model~\cite{sota} to learn features from not only the Mel-spectrogram but also the ``tempogram''~\cite{tempogram}, a feature representation of the possible tempo values of each short-time frame of a  signal.
We consider both the autocorrelation- and Fourier-tempogram proposed by Grosche \emph{et al.}~\cite{tempogram}, and experiment with both an early-fusion architecture and a late-fusion architecture.
Our experiment shows that the Mel-spectrogram only baseline attains 55.4\% song-level accuracy, and that our best model incorporating additionally tempograms achieves 60.6\% accuracy.  We find in particular salient improvement in subgenres such as
``future-house,'' ``leftfield-house-and-techno,'' and ``uplifing-trance''.

While we cannot re-distribute the audio files of the data due to copyright issues, we release our code and model checkpoints at \url{https://github.com/mir-aidj/EDM-subgenre-classifier}.   


\section{Related Work}
\label{sec:related}

\textbf{Music genre classification} is a well-researched task in the field of music information retrieval (MIR), usually formulated as a multi-class classification problem. For example, the most famous and widely-used dataset in music genre classification, the GTZAN dataset \cite{gtzan}, includes in total 1K songs covering the following 10 genres: Blues, Classical, Country, Disco, Hiphop, Jazz, Metal, Pop, Reggae and Rock.
In contrast, relatively less research has been done on music subgenre classification, which aims at distinguishing between subgenres belonging to the same ``super'' genre such as EDM and Jazz. 
We review three such existing work  below.

\textbf{EDM subgenre classification.} 
Caparrini \emph{et al.}~\cite{edm} presented a non-deep learning method for EDM subgenre classification. They compiled two different collections of EDM subgenre data from Beatport, the first with 23 subgenres and the second expanded version with 29 subgenres. They retrieved 100 songs for each subgenre from the Beatport website for both datasets.
Then, they employed 92 hand-crafted audio features and tested several non-deep learning based classifiers under 10-fold cross validation.  For the first set, the gradient tree boosting~\cite{boosting} performed the best, reaching 59.2\%  song-level classification accuracy. 
For the second set, which is more challenging due to the larger number of classes, the extremely randomised trees~\cite{ert} performed the best, reaching 48.2\% accuracy. 
They also drew two confusion matrices for both two sets, as well as the directed graph to show which subgenres were easily misclassified. 
And, they measured the importance of each feature in their tree-based classifier, finding that the top-four important features are all related to tempo.

\begin{figure*}
    \centering
        \includegraphics[width=0.25\textwidth]{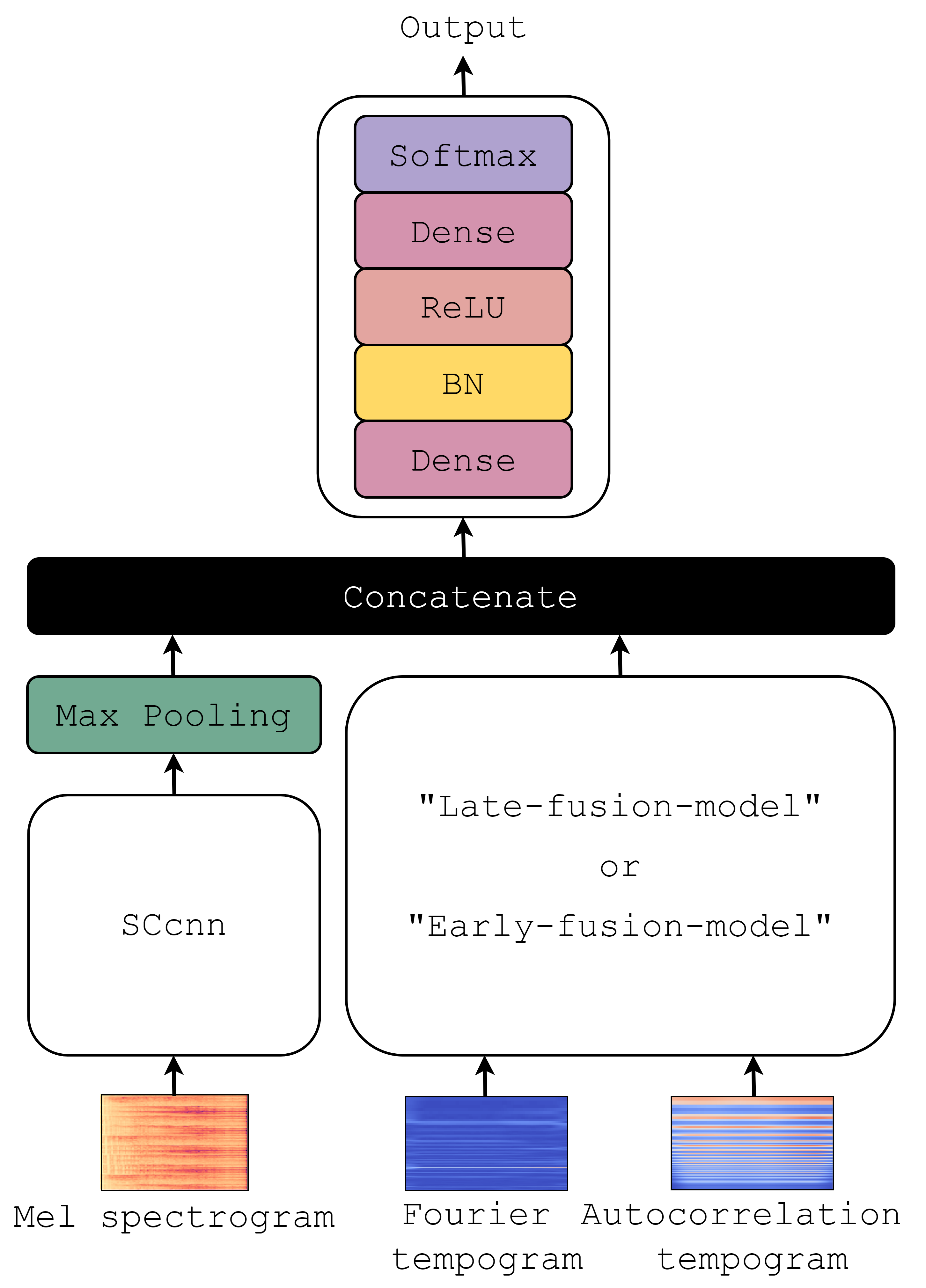}
        ~~~
        \includegraphics[width=0.27\textwidth]{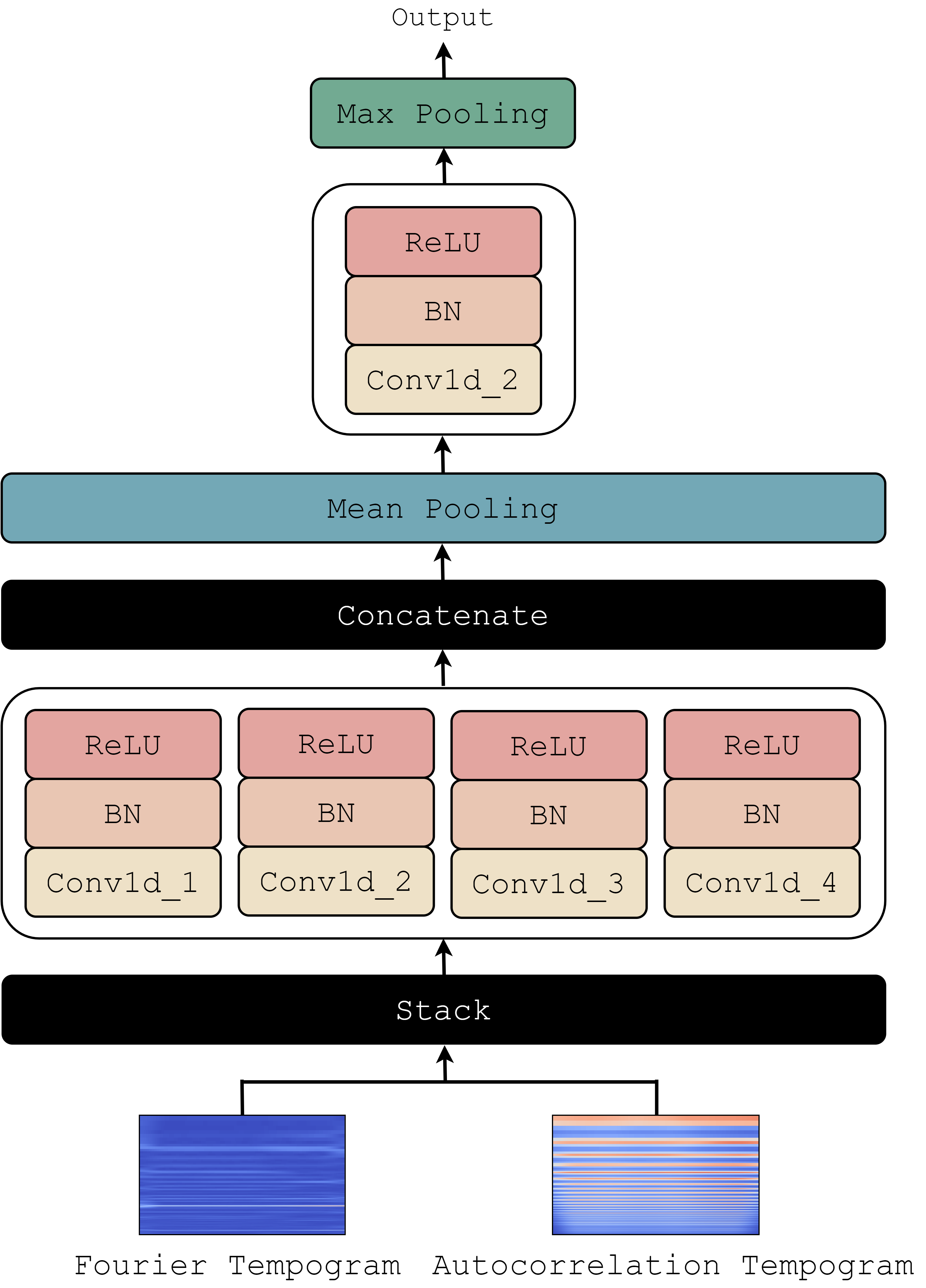}
        ~~~
        \includegraphics[width=0.41\textwidth]{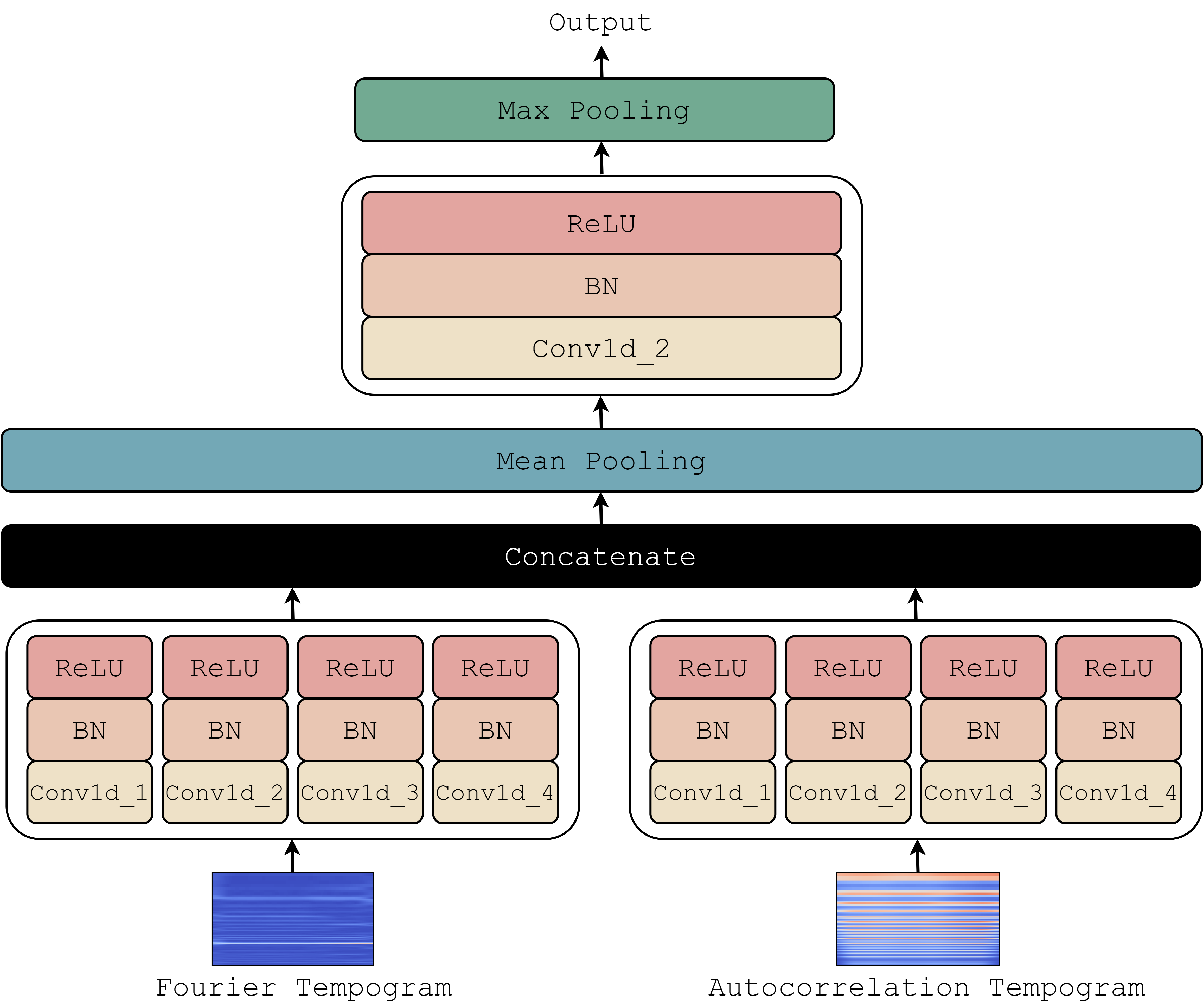} \\
        (a)\quad\quad\quad\quad\quad\quad\quad\quad\quad\quad\quad\quad\quad\quad (b)\quad\quad\quad\quad\quad\quad\quad\quad\quad\quad\quad\quad\quad\quad\quad\quad\quad\quad (c)\quad\quad\quad\quad
    \caption{System diagram of the proposed architecture for (a) employing both Mel-spectrograms and two different tempograms for music subgenre classification; (b) zoom-in of the early-fusion module for fusing the two tempograms; (c) the late-fusion variant. The `SCcnn' in (a) denotes the stack of 7 copies of feature extraction blocks (each with two 2-D convolutions) shown in Figure \ref{fig:short-chunk cnn}.}
    \label{fig:fusion-model}
\end{figure*}

\textbf{Heavy metal  subgenre classification.}
Tsatsishvili~\cite{metal} investigated subgenre classification for heavy metal music, also using non-deep learning methods. 
The author compiled a dataset of 210 songs, comprising 30 songs for each of the following 7 heavy metal subgenres: ``black,'' ``death,'' ``melodic,'' ``death,'' ``gothic,'' ``heavy,'' ``power'' and ``progressive.'' 
Half the songs were used for training and the other half for testing. 
Audio features employed in the classifiers were automatically chosen from 200 hand-crafted features with either a correlation-based feature selection method or a wrapper selection method. 
The best classification accuracy was achieved by AdaBoost~\cite{adaboost}, reaching 45.7\%.

\textbf{Jazz subgenre classification.}
Quinto \emph{et al.}~\cite{jazz} employed deep learning methods for Jazz subgenre classifcation, considering only three subgenres: ``acid-jazz,''  ``bebop,'' and ``swing/electroswing.''  
They considered a simpler multi-layer perceptron (MLP) with 1--3 layers, as well as a more sophisticated 3-layer recurrent neural network employing long short-term memory (LSTM) with 32 neurons per layer.
They employed Mel-frequency cepstral coefficients (MFCC) as the input feature. Their dataset includes 254 minutes of ``acid-jazz,'' 141 minutes ``bebop,'' and 245 minutes ``swing/electroswing.'' 60\% of them were used for training and the rest for testing. 
Finally, they got 90\% testing accuracy with the LSTM classifier, and 79\% with the MLP classifier.
The accuracy was fairly high, possibly because of the small number of classes.

Unlike music genre/subgenre classification, the \textbf{music auto-tagging} task~\cite{sota,mtat,turnbull,jynet,musicnn,harmonic,mtg,self} 
is often formulated as a multi-label classification problem, where a song can be labeled with multiple tags.  
The performance of an auto-tagging model is usually evaluated in terms of metrics such as PR-AUC and ROC-AUC \cite{spm}.
Won  \emph{et al.}~\cite{sota} experimented with a larger number of different  CNN-based models on three widely-used public datasets for this task: MagnaTagATune (MTAT)~\cite{mtat}, million song dataset (MSD)~\cite{msd}, and MTG-Jamendo \cite{mtg}, 
finding that a particular architecture called ``short-chunk CNN$+$Resnet'' performs in general the best. 
As it is straightforward to modify this architecture for multi-class classification, we employ it as the backbone architecture of our models in this work.

\section{Methods}
\label{sec:method}

\subsection{Dataset}
Following Caparrini \emph{et al.}~\cite{edm}, we build our dataset by crawling the audio previews of songs and their corresponding class labels from Beatport.
Different from their work, we consider a larger set of classes, covering the 30 EDM subgenres listed in Figure \ref{fig:boxplot}, and a much larger collection of songs, with consistently 2,500 songs per subgenre.\footnote{This Beatport dataset was initially created in one of our prior works~\cite{popmusichighlighter}.} 
We split the dataset by the ratio of 8:1:1 per subgenre to get the training, validation, and test sets. Namely, the test set contains 250 tracks per subgenre.
Every audio preview made available by Beatport is 2 minutes long. We make them consistently mono channel with 22,050 Hz sampling rate.

\subsection{Input Features: Mel-spectrogram and Tempograms}

Instead of using hand-crafted features as done by Caparrini \emph{et al.}~\cite{edm}, we employ ``raw'' features as input to our deep neural network for feature learning. The basic feature representation, also one of the most widely-used one in MIR, is the Mel-spectrogram, a time-frequency representation that is computed by applying the perceptually-motivated Mel filter bank to the spectrogram of an audio waveform. 
With \texttt{librosa}~\cite{librosa}, we compute the Mel-spectrogram using a Hamming window of 2,048-sample long and 512-sample hop length for the short-time Fourier transform (STFT), and 128 Mel filters.

We also employ the tempogram~\cite{tempogram} as input feature, a ``time-tempo'' representation that contains local tempo information for each frame of an audio signal. Grosche \emph{et al.}~\cite{tempogram} proposed two types of tempogram: the \emph{Fourier tempogram} and the \emph{autocorrelation tempogram}. 
The former converts frequency (Hz) to tempo (beat-per-minute; BPM), emphasizing the harmonics, while the latter converts time-lag (seconds) to tempo, emphasizing instead the subharmonics. We consider both here.
The \emph{Fourier tempogram}\footnote{\url{https://www.audiolabs-erlangen.de/resources/MIR/FMP/C6/C6S2_TempogramFourier.html}} is computed by firstly estimating from the audio waveform a ``novelty curve'' indicating note onset candidates, and then computing the Fourier representation of the novelty curve (not the original audio waveform) using STFT. 
The resulting representation is assumed to capture local periodic patterns of the input signal.
Its frequency axis is finally mapped to a BPM axis. 
The \emph{autocorrelation tempogram},\footnote{\url{https://www.audiolabs-erlangen.de/resources/MIR/FMP/C6/C6S2_TempogramAutocorrelation.html}} on the other hand, is computed by calculating the local autocorrelation function for different time lags from the also the novelty curve, and then converting the time-lag to a linear BPM axis using interpolation and resampling. 
We also employ \texttt{librosa}~\cite{librosa} to compute these two variants of the tempogram, with 512-sample hop size and a Hamming window of 2,048 samples. 
Figure \ref{fig:spec_temp_compare} provides examples of the Fourier tempogram and autocorrelation tempogram of songs in our dataset.\footnote{Grosche \emph{et al}~\cite{tempogram} also proposed the ``cyclic tempogram,'' where tempi differing by a power of two are identified, but we do not use it here.}

We use two different length per song to compute the Mel-spectrograms: the 30-second segment from 15s to 45s of a song, or the whole two minutes. For the tempograms, we use only the segment a song from 15s to 45s for simplicity.
The size of a Fourier tempogram and an autocorrelation  tempogram would be 193$\times$1,293 (BPM$\times$time) and 384$\times$1,292,
respectively.
Audio is fed to our models using ``chunks'' (see below) of such fixed-length Mel-spectrograms and tempograms. Every feature dimension is zscore-normalized.

\subsection{Short-chunk CNN with Resnet}
We employ the short-chunk CNN$+$Resnet as the backbone architecture of our models, as it has been shown to outperform competing models in music auto-tagging across different benchmark datasets~\cite{sota}. 
As depicted in Figure \ref{fig:short-chunk cnn}, this model contains 7 copies of feature extraction layers, each comprising two 2-D convolutional layers, two batch normalization layers (after convolution), and one ReLU activate layer in between convolutions. The output of this stack of layers goes through max pooling and then two dense layers and a final softmax layer for classification. The name ``short-chunk'' stems from the fact that the model is designed to take as input short segments of the Mel-spectrogram of the size 128$\times$200, which in other words divides our Mel-spectrogram into 25 chunks (neglecting the last 168 frames). We similarly divide the tempograms into 25 chunks. Following \cite{sota}, we assume that the subgenre label of each chunk is the same as the the subgenre label of the song the chunk comes from.

\begin{table}
    \caption{Chunk- and song-level testing accuracy of 30-class EDM subgenre classification of the evaluated short-chunk CNN-based models; the first two models use only the Mel-spectrograms, while the last two use both the Mel-spectrograms and tempograms}
    \centering
    \begin{tabular}{l|cc}
        \toprule 
         & \textbf{chunk-level} &  \textbf{song-level} \\
        \midrule
        Mel-spectrogram only (30 sec) \cite{sota}      & 46.1\% & 50.4\% \\
        Mel-spectrogram only (120 sec) \cite{sota}      & 46.1\% & 55.4\% \\
        \midrule
        Fourier tempogram only (30 sec) & 32.0\% & 34.9\%\\ 
        autocorrelation tempogram only  (30 sec) & 28.3\% & 31.2\%\\
        \midrule
        early-fusion & 53.4\% & 60.3\% \\
        late-fusion & 53.3\% & 60.6\%\\
        \bottomrule
    \end{tabular}
    \label{fig:accuracy_chunk_song}
\end{table}

\subsection{Proposed Fusion Models}

Figure \ref{fig:fusion-model}(a) shows the architecture we propose to integrate the (chunk-level) Mel-spectrogram, Fourier tempogram, and autocorrelation tempogram of a song for  classification.
We firstly fuse the two tempograms into a combined representation, and then concatenate it with the output of the feature extraction blocks of the short-chunk CNN branch that deals with the Mel-spectrogram. After feature concatenation, we use the same classification block of the short-chunk CNN.

Figures \ref{fig:fusion-model}(b) and (c) depict the two fusion strategies considered in our work to combine the two tempograms. 
They both use four parallel 1-D convolutional layers with different kernel sizes (3, 3, 5, 5) and strides (2, 3, 3, 5) for feature extraction, a design that is inspired by the work of Pons \emph{et al.}~\cite{b6}. The outputs of the 1-D convolution layers are concatenated, mean-pooled, futher processed with a 2-D convolution layer, and finally max-pooled to yield a combined representation. 
The \emph{early-fusion} variant (Figure \ref{fig:fusion-model}(b)) combines the two tempograms at the very beginning with feature-wise concatenation, while the \emph{late-fusion} variant (Figure \ref{fig:fusion-model}(c)) combines them after the 1-D convolutional layers.

\begin{figure}
    \centering              \includegraphics[width=0.484\textwidth]{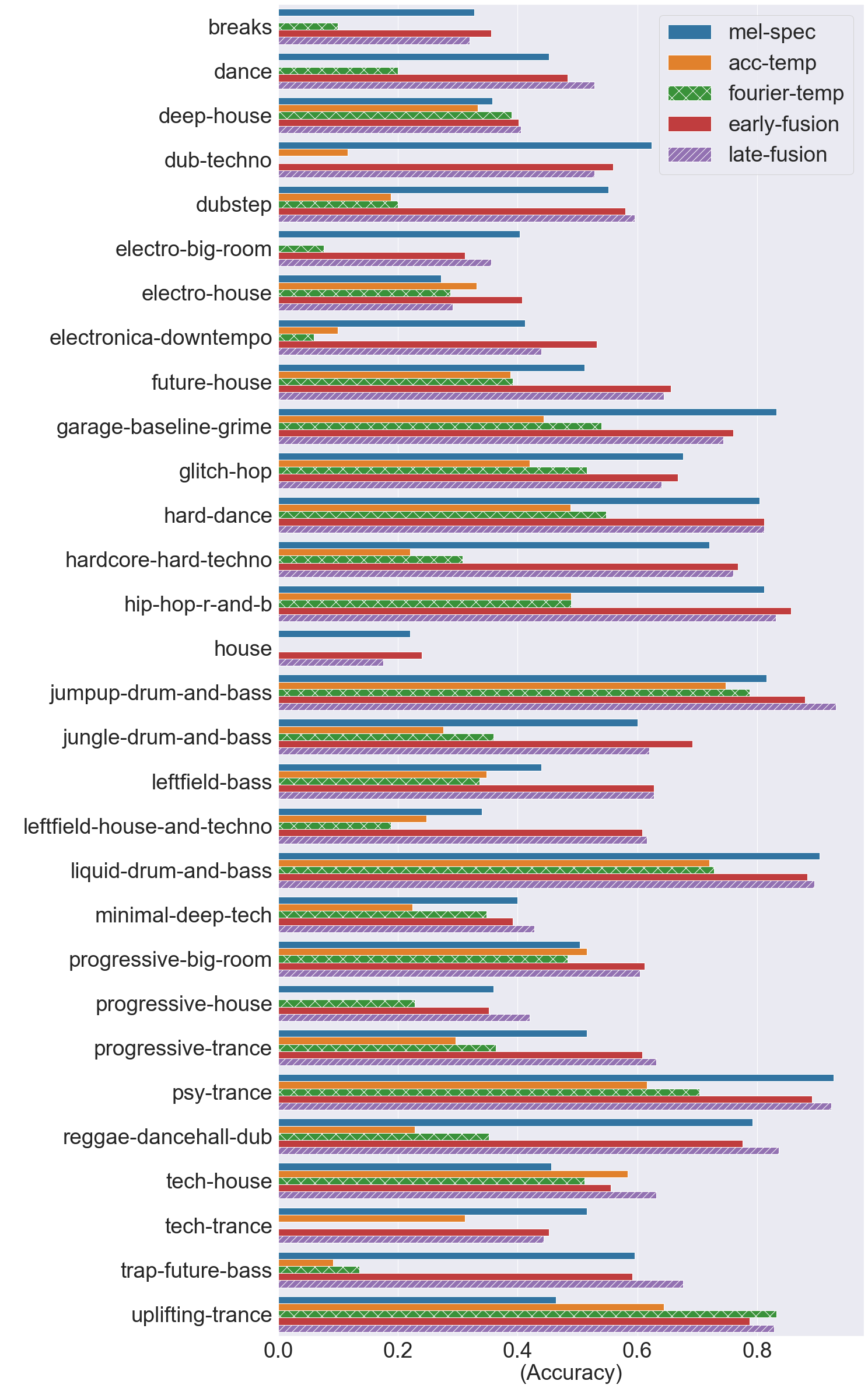}
    \caption{Per subgenre testing classification accuracy of different models.}
    \label{fig:barplot_acc_comparison}
\end{figure}


\section{Experiments}
\label{sec:exp}

We train the following deep models by using cross-entropy as the loss function and Adam as the optimizer. 
\begin{itemize}
    \item \emph{Mel-spectrogram only}: the short-chunk CNN$+$Resnet baseline \cite{sota} shown in Figure \ref{fig:short-chunk cnn}.
    \item \emph{Proposed early- and late-fusion}: the proposed models that use both the Mel-spectrograms and the two tempograms, shown in Figure \ref{fig:fusion-model}.
    \item \emph{Fourier tempogram or autocorrelation tempogram only}: the ablated variants that use only one of the tempograms, without using the Mel-spectrograms. This is done with a simplified version of the fusion models, with four 1-D convolutional layers and a 2-D convolutional layer for feature extraction, and two dense layers for classification. 
\end{itemize}
We set the batch size to 256, and train the models for 200 epochs with learning rate $0.005$. We use 50\% dropout before the last dense layer and after the ReLU layer.

We report the \textbf{chunk-level} classification accuracy by taking the class with the highest softmax-ed value as the prediction result. 
In addition, we also report the \textbf{song-level} accuracy by a majority voting mechanism over the prediction results of the short chunks of a song.

Table \ref{fig:accuracy_chunk_song} shows that early-fusion and late-fusion models are both better than Mel-spectrogram-input model in both chunk-level and song-level accuracy. The late-fusion model performs overall the best, reaching 60.6\% song-level testing accuracy for our 30-class classification setting, which is much higher than the 48.2\% accuracy obtained by Caparrini \emph{et al.}~\cite{edm} for 29-class classification using a non-deep learning method.This also stands as a 
$+$5 to $+$10\% performance gain relative to the Mel-spectrogram only baseline, empirically validating the benefit of using tempo-related features for EDM subgenre classification. 
Table \ref{fig:accuracy_chunk_song} also shows that the Fourier tempogram only model outperforms the autocorrelation tempogram counterpart.

\begin{figure}[t]
    \centering
        \includegraphics[width=0.484\textwidth]{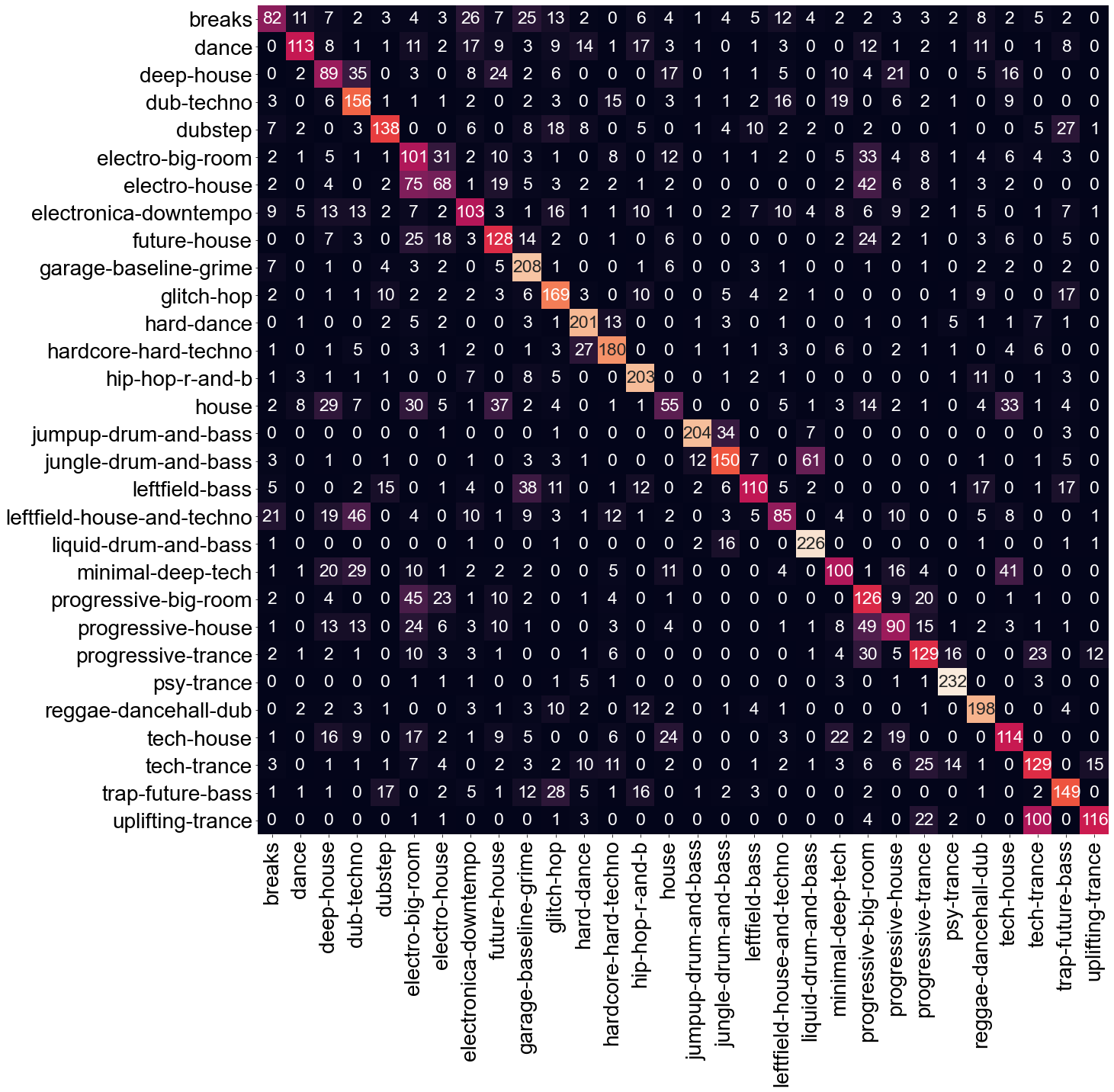}\\
        \vspace{1mm}
        \small{(a) Confusion table of the short-chunk CNN baseline (120 sec)}
        \\
        \vspace{3mm}
        \includegraphics[width=0.484\textwidth]{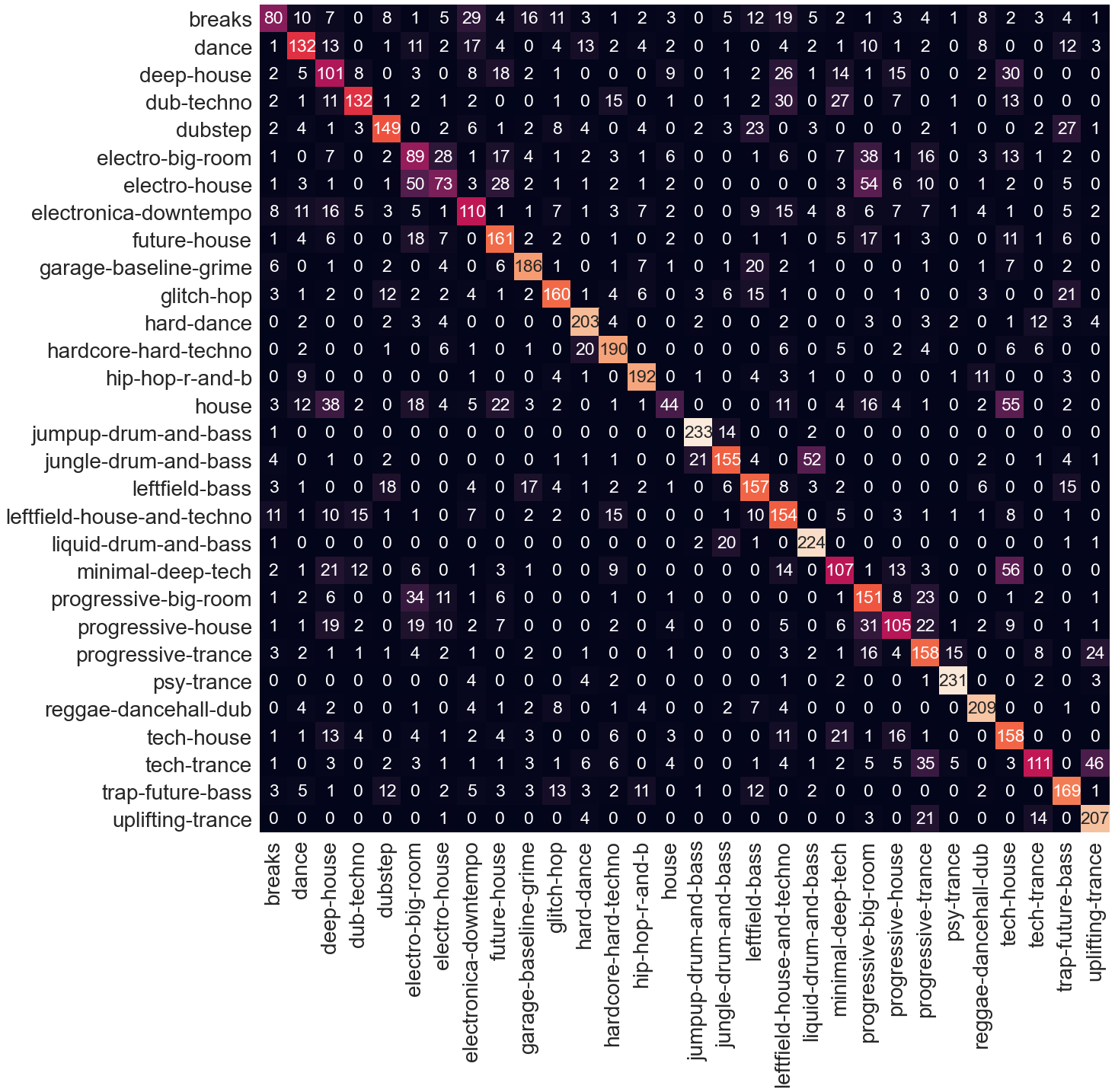} \\
        \vspace{1mm}
        \small{(b) Confusion table of the proposed late-fusion model}
    \caption{The confusion matrices of the testing result of two evaluated models. Each row shows whether the songs from a subgenre are misclassified to other subgenres; those on the diagonal are correctly classified. As our test set contains consistently 250 songs per subgenre, each row sums up to 250.}
    %
    \label{fig:cm-short-chunk-cnn}
\end{figure}

Figure \ref{fig:barplot_acc_comparison} shows the per-genre result of these five models.
We can see salient performance improvement for genres such as
``future-house,'' ``leftfield-house-and-techno'' and ``uplifing-trance''
for the proposed fusion models than the baseline Mel-spectrogram only model.

Figure \ref{fig:cm-short-chunk-cnn} shows the 30$\times$30 confusion matrices of the baseline model and the proposed late-fusion model. We can see that the propose model still gets confused for some pairs of subgenres, but less seriously so than the baseline model.

\begin{figure*}
    \centering
        \includegraphics[width=0.484\textwidth]{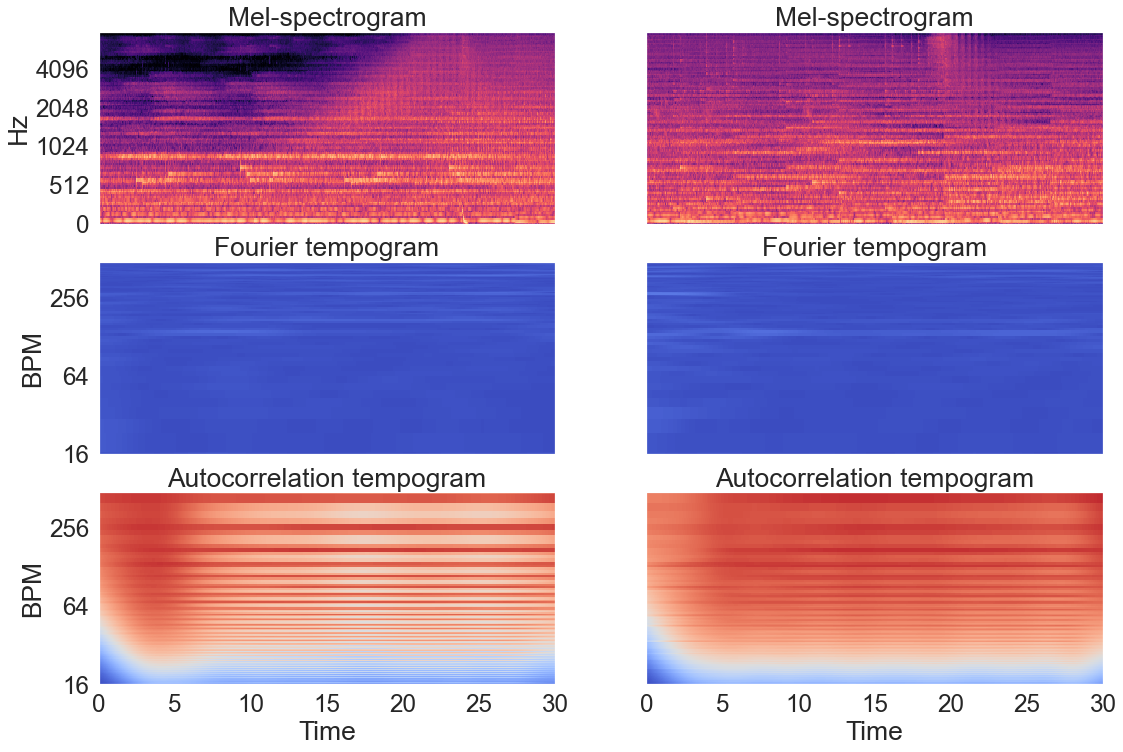}
        \includegraphics[width=0.484\textwidth]{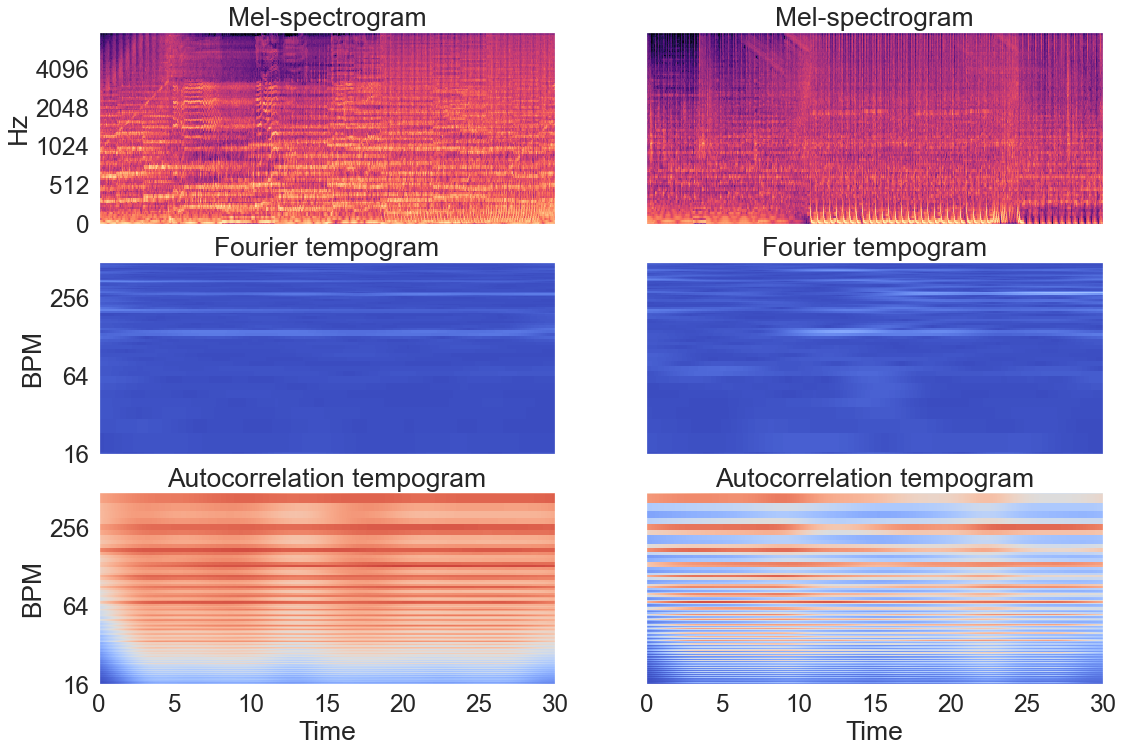} \\
        \quad\quad (a) ~~\quad\quad\quad\quad\quad\quad\quad\quad\quad\quad (b) ~~\quad\quad\quad\quad\quad\quad\quad\quad\quad\quad\quad (c) \quad\quad\quad\quad\quad\quad\quad\quad\quad\quad\quad (d)
    \caption{The Mel-spectrogram, Fourier tempogram and autocorrelation tempogram for (a) an ``uplifting-trance'' song misclassified as  ``tech-trance'' by the baseline model but correctly classified by our late-fusion model; (b) an ``uplifting-trance'' song correctly classified by both the baseline model and our fusion model; (c) another ``uplifting-trance'' song misclassified as ``tech-trance'' by the baseline model but correctly classified by our late-fusion model; (d) a ``tech-trance'' correctly classified by both the baseline model and our fusion model.}
    \label{fig:spec_temp_compare}
\end{figure*}

Finally, Figure \ref{fig:spec_temp_compare} visualizes the Mel-spectrograms and tempograms of four songs from our test set.
The songs shown in Figures \ref{fig:spec_temp_compare}(a) and (b) are both 
``uplifting-trance'' songs. They can both be correctly classified by our late-fusion model, but the first song would be wrongly recognized as a 
``tech-trance'' song by the baseline model. We  see that these two songs seem to have fairly different patterns in the Mel-spectrograms, but similar patterns in the autocorrelation tempograms. 
The baseline model does not work well for this song, possibly because it only has access to the Mel-spectrograms. 
On the other hand, the songs shown in Figures \ref{fig:spec_temp_compare}(c) and (d) are an 
``uplifing-trance'' song and a 
``tech-trance'' song, respectively. Again, they can both be correctly classified by our late-fusion model, but the first song in this pair (i.e. (c)) would be wrongly regarded as a 
``tech-trance'' song by the baseline model.
We see these two songs, despite they are associated with different subgenres, have some similar local patterns in their Mel-spectrograms, which might have caused confusion for the baseline model.


\section{Conclusion}
\label{sec:conclusion}
In the paper, we have presented a deep learning approach for EDM subgenre  classification, achieving 60.6\% testing accuracy for 30-class classification when using not only the Mel-spectrograms but also the tempograms as input feature.
We found that the proposed late-fusion model is 10\% more accurate than a Mel-spectrogram only deep learning baseline model.
This may suggest that, for a fine-grained classification problem such as EDM subgenre classification, it is beneficial to consider multiple inputs for feature learning.

We have a few ideas for future extension of this work. First, we would like to explore different architectures to process the tempograms, such as the use of recurrent layers or temporal convolutional layers \cite{tcn}.
Second, to incorporate more features as input, including other mid-level features \cite{deeprhythm}  
as well as high-level features \cite{musicnn,popmusichighlighter,hit}. Third, to employ other feature fusing strategies. And finally, to employ metric-based learning method \cite{tag_metric}, which is shown to be  promising in recent work on music auto-tagging.

\end{document}